\documentclass[amsmath,amssymb,twocolumn,prl]{revtex4}
\usepackage{color}
\usepackage[]{epsfig}
\newcommand{\revise}[1]{{\color{black} {#1}}}
\newcommand{\gdot}{\dot{\gamma}}
\renewcommand{\P}{\Pi}
\newcommand{\p}{\bar\Pi}
\newcommand{\Pe}{{\rm Pe}}
\newcommand{\Pen}{{\rm Pe_0}}

\begin{document}

{\bf Comment on `Constant stress and pressure 
rheology of colloidal suspensions'.}

In a recent Letter, Wang and Brady (WB)~\cite{WB} analyse the rheology of 
Brownian hard spheres using \revise{constant stress 
and pressure Brownian dynamics} 
simulations. The main 
observable is the shear viscosity, $\eta(\gdot, \P)$, \revise{expressed as
a function of the shear rate} $\gdot$ and 
adimensional pressure $\p=\P a^3/(k_BT)$,
where $\P$ is the pressure, $k_B T$ the thermal 
energy, and $a$ the average particle diameter. 
The central conclusion is the discovery of a ``universal viscosity 
divergence''~\cite{WB}, 
\begin{equation}
\eta \sim [\phi_m(\p)-\phi]^{-\gamma}, \ \ \ 
\gamma \simeq 2,
\label{wbeq}
\end{equation}
where $\phi$ is the volume fraction and $\phi_m(\p)$ a 
pressure-dependent critical density. 
WB argue that Eq.~(\ref{wbeq}) is valid for all $\p$, unifying
the viscosity divergence of both thermal 
and athermal assemblies of hard spheres. Assuming that 
(\ref{wbeq}) describes the jamming transition of athermal hard spheres 
(at $\p \to \infty$), they conclude that the same physics
must control the rheology at finite $\p$. In this view, pressure 
only changes $\phi_m(\p)$, which interpolates between 
the jamming density $\phi_J$ 
for $\p\to\infty$ and the glass transition density
at low $\p$~\cite{WB}, implying that Eq.~(\ref{wbeq})
is valid also for the glass transition of Brownian hard 
spheres at equilibrium.
 
\revise{We show that these conclusions are not valid
and provide the appropriate perspective to interpret
WB's results.} We argue that the reported universality 
stems from exploring a single rheological regime
where the \revise{hard sphere thermal glass} is non-linearly sheared 
beyond yield, which directly explains the universal
value $\gamma \approx 2$ in terms of the known hard sphere glass rheology, 
with no connection to the Newtonian
regimes of either the colloidal glass transition or the 
granular jamming transition.

\begin{figure}
\begin{center}
\psfig{file=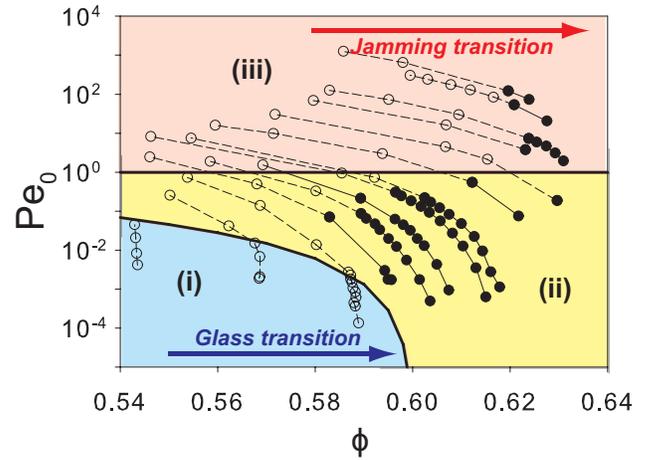,width=8.cm}
\end{center}
\caption{Using adimensional shear rates, we organize the data of WB
measured along constant-$\p$ paths (dashed lines) 
into three regimes. Low viscosity data, $\eta<40$, are shown
as empty circles.  (i) Newtonian thermal; 
(ii) shear-thinning thermal; 
(iii) Newtonian athermal. 
Filled symbols indicate data that 
are described by Eq.~(\ref{wbeq}), having $\eta > 40$. 
As most of these data lie in (ii), 
Eq.~(\ref{wbeq}) follows from the known rheology of 
the thermal hard sphere glass. The transition line between (ii) and (iii)  
is at $\Pen = 1$; the one between (i) and (ii) ($\Pe=1$) 
is determined by adjusting the viscosity model of \cite{IBS} to WB's data.}
\label{fig}
\end{figure}

To assess the role of thermal fluctuations in hard sphere rheology, 
it is useful to consider not only the reduced pressure $\p$ but also 
the timescales associated with thermal fluctuations~\cite{IBS}.
We consider two P\'eclet numbers: 
$\Pen = \gdot a^2 / d_0$ and $\Pe = \gdot a^2 / d$,
where $d(\phi)$ is the single particle diffusion 
coefficient at $\gdot=0$, and $d_0$ its dilute limit. \revise{Only 
$\Pe$ is considered in \cite{WB}, but}  
because $\Pen < \Pe$, three different regimes exist, which 
we use in Fig.~\ref{fig} to organise the data of WB. 

{(i)} $\Pen < \Pe \ll 1$: The shear flow is slower than equilibrium 
relaxation ($\Pe \ll 1$). 
In this equilibrium regime, the viscosity is Newtonian,
$\eta = \eta_{T}(\p)$~\cite{IBS}. It depends only on $\p$ and so cannot 
be varied in the constant-$\p$ paths of WB. By construction, 
this approach cannot follow the rapid growth of the Newtonian 
viscosity of Brownian hard spheres on approaching the glass transition,
which is indeed known~\cite{thermalHS} to differ from Eq.~(\ref{wbeq}).
 
{(ii)} $\Pen \ll 1 \ll \Pe$: This is
the shear-thinning  regime ($1 \ll \Pe$)  of the 
thermal ($\Pen \ll 1$)
hard sphere glass.
Most data described by~(\ref{wbeq}) are in this regime (Fig.~\ref{fig}). 
WB's universal collapse
therefore describes the divergence of a 
{\it non-Newtonian} viscosity.
These constant-$\p$ observations can be accounted for using
known laws for the rheology of amorphous solids \revise{(including 
Brownian hard spheres).} 
We assume Herschel-Bulkley rheology for both $\sigma$ 
and $\p$:
$\p(\phi,\gdot) = \p_y(\phi) + b(\phi) \gdot^{n}$, where $\p_y(\phi)$, 
the inverse function of $\phi_m(\p)$, is the pressure at yield. 
Expanding to linear order in density change at constant $\p$ gives
$\phi_m(\p) - \phi(\gdot,\p) \sim \gdot^n$, 
thus $\eta = \sigma / \gdot \sim \sigma_y/\gdot \sim [\phi_m(\p) 
- \phi]^{-1/n} $, 
showing that $\gamma = 1/n$. Numerical results are well described 
by $n \simeq 0.5$~\cite{IBS}, in good agreement with the value 
$\gamma\simeq 2$
reported by WB. Data in the inset of their Fig.~3a are also 
consistent with the known yield pressure divergence near 
jamming~\cite{rintoul}, 
from which we predict $\phi_m (\p) \sim \phi_J - c/\p$ at large $\p$.

{(iii)}  $1 \ll \Pen  < \Pe$: Shear flow dominates even 
local thermal motion ($1 \ll \Pen$), so this regime is effectively athermal
and the only one where jamming physics could be explored. 
The viscosity is Newtonian, 
$\eta=\eta_0(\phi)$ and only depends on $\phi$ or, in WB's setup, 
$\sigma/\P$. As shown in Fig.~\ref{fig} the data 
of WB are too sparse in this regime to assess the functional
form of $\eta_{0}(\phi)$. Recent results show that Eq.~(\ref{wbeq})
is again not valid asymptotically here~\cite{takeshi}. \\

A. Ikeda$^1$, L. Berthier$^2$ and P. Sollich$^3$ \\

$^1$Fukui Institute for Fundamental Chemistry, Kyoto University, Kyoto, Japan

$^2$Laboratoire Charles Coulomb, UMR 5221 CNRS-Universit\'e 
de Montpellier, Montpellier, France 

$^3$King's College London, Department of Mathematics, Strand, 
London WC2R 2LS, UK

\end{document}